\documentclass[aps,pra,twocolumn,showpacs,preprintnumbers,amsmath,amsfonts,amssymb,footinbib,floatfix]{revtex4-2}
\usepackage{mathtools}

\usepackage{braket}
\usepackage{graphicx,epsfig}
\usepackage{bm}
\usepackage{dcolumn}
\usepackage{xcolor}
\usepackage{lipsum}
\usepackage{float}
\usepackage[breaklinks=true,colorlinks,citecolor=blue,linkcolor=blue,urlcolor=blue]{hyperref}

\usepackage[capitalize]{cleveref}

\def\noi{\noindent}
\def\bc{\begin{center}}
	\def\ec{\end{center}}
\newcommand{\bea}{\begin{equation}}
	\newcommand{\eea}{\end{equation}\noi}
\newcommand{\ber}{\begin{eqnarray}}
	\newcommand{\eer}{\end{eqnarray}\noi}

\begin{document}	
	
\title{Many boson quantum Szilard engine for fractional power law potential}
	
\author{Najirul Islam}\email{nislam@tifrh.res.in}

\affiliation{Tata Institute of Fundamental Research, Hyderabad, Telangana 500046, India}

\date{\today}
	
\begin{abstract}
In this article, we have realized the quantum Szilard engine (QZE) for non-interacting bosons. We have adopted the Bose-Einstein statistics for this purpose. We have considered fractional power law potential for this purpose and have used the artifact of the quantization of energy. We have calculated the work and the efficiency for non-interacting bosons in fractional power potential. We have shown the dependence of the number of particles for the work and the efficiency. We also have realized the QZE for a single-particle in a Morse potential revealing how the depth of the potential impacts both work and efficiency. Furthermore, we have examined the influence of temperature and the anharmonicity parameter on the work. Finally, we have conducted a comparative analysis, considering both non-interacting bosons in a fractional power law potential and a single-particle in a Morse potential under harmonic approximation conditions.
\end{abstract}
	
	
\maketitle
	
	
\section{Introduction}
The concept of Maxwell's demon establishes the relationship between thermodynamics and information in theoretical physics \cite{leff2002maxwell}. In this thought experiment a demon sitting in the position of the wall, selects molecules of fast velocity to go one side and molecules of slow velocity on the other side of the wall. This creates a temperature difference between the two sides of the particles and hence a decrease in the entropy of the system without doing any work. Thus Maxwell's demon violates the second law of thermodynamics (SLT) \cite{leff2002maxwell,maxwell1990scientific}.
	
The first explanation of this violation of the SLT was brought by Szilard in the form of classical analysis. He made a connection between entropy and information in the form of the Szilard engine \cite{leff2002maxwell,szilard1929entropieverminderung}. The idea of the Szilard engine consists of a classical particle contained in a container that can be divided by a movable barrier. The movable barrier is coupled to a heat bath. The demon measures the position of the particle on either side of the wall and records it. Now absorbing heat from the bath, the gas expands isothermally. Thus the Szilard engine transforms heat isothermally into work of amount $k_B T\log{(2)}$ where, $T$ is the temperature of the bath \cite{thomas2019quantum,leff2002maxwell,szilard1929entropieverminderung,maxwell1990scientific}. Thus the spooky action of the demon results in a decrease of entropy of the bath of amount $k_B \log{(2)}$ and thus violates the SLT \cite{thomas2019quantum}. The contradiction of the violation of the SLT was resolved by Szilard accounting the work associated with the information of the demon \cite{bengtsson2018quantum,thomas2019quantum,szilard1929entropieverminderung,leff2002maxwell}. The erasure of one bit of information costs at least an entropy of amount $k_B \log{(2)}$ by the Landauer’s erasure principle \cite{thomas2019quantum,bengtsson2018quantum,landauer1961irreversibility,leff2002maxwell,bennett1982thermodynamics,sagawa2008second}. This was verified experimentally for single classical particle \cite{bengtsson2018quantum,toyabe2010experimental,koski2015chip,roldan2014universal}.
	
\begin{figure}
\includegraphics[width=1.00 \linewidth]{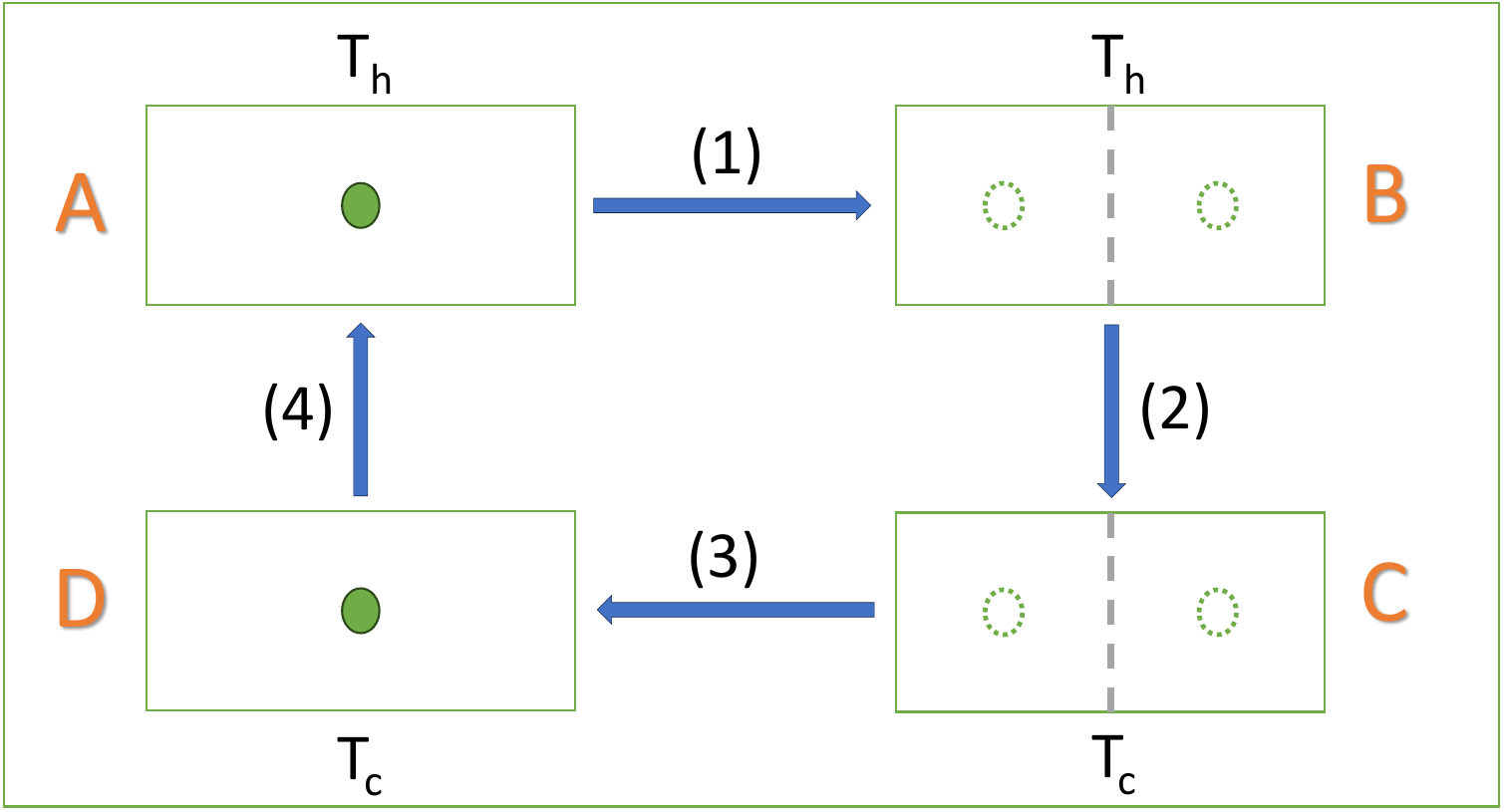}
\caption{Schematic view of a single-particle QZE. The green circles represent the particle without a barrier in the middle. The green dashed circles represent the uncertainty of the particle (left or right of the barrier) after introducing the barrier in the middle. The figure shows the four stages of the Stirling cycle. Processes (1) and (3) are the isothermal insertion and isothermal removal while processes (2) and (4) are isochoric. The four boxes represent the potential well. While the boxes A and B are coupled to bath of temperature $T_h$, the boxes C and D are coupled to a bath of temperature $T_c$ and $T_h>T_c$.}
\label{fig:1}
\end{figure}
	
Still, we know very less about how the connection between work, entropy, and information varies for many-particle systems and for different potentials. In the classical limit, the insertion and removal of the barrier do not cost an extra amount of work \cite{sur2023quantum,kim2011quantum}. However this is not true in the quantum regime because the insertion of the barrier shifts the energy eigenvalues \cite{thomas2019quantum,bengtsson2018quantum,moore2012frontiers,gea2005quantum,bender2005unusual,gea2002splitting,belloni2014infinite}.
	
An ideal Stirling cycle consists of isothermal heat addition and isothermal heat removal processes and two isochoric processes \cite{romanelli2017alternative}. The transition of the QZE from the classical Szilard engine comes from treating the working medium (particle) quantum mechanically. Thus, a particle in a potential well will play the role of the container \cite{zurek1984maxwell}. work will be described by virtue of the quantization of energy. Thus, the partition function can describe the thermodynamic behavior of the system \cite{bengtsson2018quantum,zurek1984maxwell}. Additionally, the seminal work by Rosen on the transition from quantum mechanics to classical mechanics is given by the order of the mass of the particle \cite{rosen1964relation,rosen1986quantum}. The standard mass of this transition point is given by the Planck mass as $m_p=\big(\frac{\hbar c}{G}\big)^{1/2}=2.17\times10^{-8}~$kg, where $G$ is the gravitational constant \cite{adler2010six,rosen1964relation,rosen1986quantum}. A particle with a mass much smaller than $m_p$ obeys the Schrodinger wave equation while a particle of mass of the order of $m_p$ or higher values obeys the `non-linear classical Schrodinger equation' where superposition principle doesn't hold \cite{rosen1986quantum}. 
	
The advent of technologies to control the system of quantum particles has eased the study of quantum thermodynamics. Since at the nanoscale, the quantum effects dominate, the study of quantum heat engines is prominent to extract better work and improved efficiency beyond classical counterpart \cite{purkait2022performance,huang2014quantum,raja2021finite,chen2016classical,cruz2023quantum}. In the classical Szilard engine the internal energy remains conserved throughout all the thermodynamic processes while in the quantum model, it changes during insertion and removal of the barrier \cite{bracken2014quantum,zurek1984maxwell,gea2002splitting,bender2005unusual,gea2005quantum}. 
	
Szilard's original idea about the Szilard engine was considering a single classical particle \cite{szilard1929entropieverminderung}. Kim et al. showed first that the work extracted and the efficiency of two interacting bosons or two interacting fermions is greater than a single classical particle due to the indistinguishability of quantum particles \cite{kim2011quantum}. However, the amount of work extracted depends on the particle statistics \cite{kim2011quantum}. The indistinguishability contains more information than the distinguishability nature of the particles \cite{kim2011quantum}. Since the amount of indistinguishability increases with the increasing number of indistinguishable particles, leads to a greater amount of extraction of work and greater efficiency \cite{kim2011quantum}. 
	
It is hard to treat many indistinguishable particles to realize a QZE analytically. The Bose-Einstein statistics are effective in the low-temperature regime. The number of available states increases in the high-temperature limit. The thermal fluctuations wash out the indistinguishability of the particles since then the particles become distinguishable by occupying different states \cite{kim2011quantum}. As mentioned earlier the amount of work to be extracted depends on the lack of information (in this case the indistinguishability of the particles and hence their location on either side of the barrier), the work and efficiency are restricted to the classical single-particle limit \cite{kim2011quantum}.
	
In this article, we have shown that the work and the efficiency of a QZE for the non-interacting bosons increase with the number of particles for a suitable range of the temperatures of the heat baths. By the word `suitable range of the temperature', we mean the behavior of the chemical potentials before and after inserting the barrier. We have shown that the efficiency increases with the increase in the particle number for non-interacting bosons in fractional power-law potential. The maximum efficiency is however restricted by the Carnot efficiency \cite{kim2011quantum,huang2014quantum,zurek1984maxwell,sur2023quantum}.  Since the non-interacting bosons partition function has a dependency on the chemical potential explicitly, the extraction of positive non-zero work and hence efficiency is only possible for the converging nature of the grand partition function. In this article, we have also calculated the work and the efficiency of a QZE for a single-particle in a Morse potential for different temperatures of the hot and the cold bath respectively. 
\section{Two non-interacting particles in a quantum harmonic oscillator}
\begin{figure}
\includegraphics[width=1.00 \linewidth]{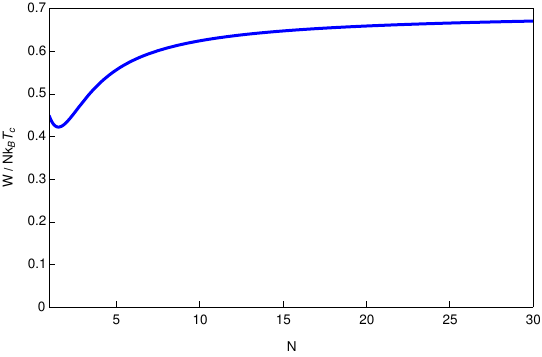}
\caption{Plot of the average work Eqn. (\ref{eqn:7}) with respect to the non-interacting particle number $N$ in a harmonic potential. Here we have used $m=19.11\times10^{-11}$kg, $\hbar=1.0545\times10^{-34}$J.s, $\omega=100$ GHz, $k_B=1.3806\times10^{-23}$J.$K^{-1}$, $T_h=200$K, and $T_c=100$K.}
\label{fig:2}
\end{figure}

\begin{figure}
\includegraphics[width=1.00 \linewidth]{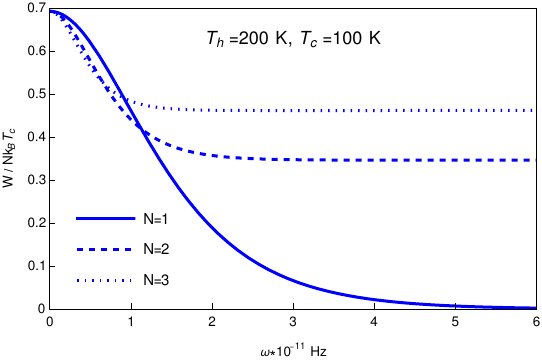}
\caption{Plot of the average work Eqn. (\ref{eqn:7}) for $N=1, 2, \text{and}~ 3$ non-interacting particles in a harmonic potential. Here we have used $m=19.11\times10^{-11}$kg, $\hbar=1.0545\times10^{-34}$J.s, $k_B=1.3806\times10^{-23}$J.$K^{-1}$, $T_h=200$K, and $T_c=100$K.}
\label{fig:3}
\end{figure}

The discrete energy spectrum of a single-particle of mass $m$ in a quantum harmonic oscillator is given by
\begin{eqnarray}\label{eqn:1}
E_n=\Big[n+\frac{1}{2}\Big]\hbar\omega,~~~~~~~~~~n=0,1,2... \infty
\end{eqnarray}
For two non-interacting distinguishable particles `A' and `B' of mass $m_A$ and $m_B$ in a harmonic oscillator, the composite energy eigenvalues and the wave function are given by \cite{griffiths2018introduction}
\begin{align}\label{eqn:2}
E_{n_A,n_B}=\Big[n_A+\frac{1}{2}\Big]\hbar\omega_A+\Big[n_B+\frac{1}{2}\Big]\hbar\omega_B\\
\text{and}~~~~~~~~~~~~~~~~~~~~~~~~~~~~~~~~~~~~~~~~~~~~~~~~~~~~~\nonumber\\
\Psi_{n_A,n_B}(x_A,x_B)=\psi_{n_A}(x_A)\psi_{n_B}(x_B).~~~~~\label{eqn:3}
\end{align}
The partition function of the combined system at temperature $T~$K can be written as 
\begin{eqnarray}
Z&=&z_A\times z_B\nonumber\\
&=&\sum_{n_A=0}^{\infty}e^{-\beta \hbar\omega_A\Big(n_A+\frac{1}{2}\Big)}\sum_{n_B=0}^{\infty}e^{-\beta \hbar\omega_B\Big(n_B+\frac{1}{2}\Big)}\label{eqn:4}
\end{eqnarray}
where $\beta=1/k_BT$. If the particles are non-interacting and indistinguishable i.e. if we don't have knowledge of which particle is in which energy levels then for $N$ indistinguishable particles, the combined partition function can be written as 
\begin{eqnarray}\label{eqn:5}
Z=\frac{z^N}{N!}.
\end{eqnarray} 
For a single-particle, there is double degeneracy in the quantized energy levels if a single barrier is inserted isothermally \cite{thomas2019quantum,zurek1984maxwell,kim2011quantum}. For $N$ non-interacting indistinguishable particles there will be $2^N$ degeneracy in the combined partition function when in boxes B and C. Suppose $N$ non-interacting particles are in a harmonic potential symmetric about the origin. A barrier is inserted in the potential at $x=0$ isothermally at stage (1). The barrier is isothermally removed at stage (3). The combined partition functions of $N$ non-interacting particles can be written as
\begin{subequations}
\begin{eqnarray}\label{eqn:6}
Z^N_A&=&\sum_{n=1}^{\infty}\Big[e^{-\frac{\hbar\omega\big(n+\frac{1}{2}\big)}{k_B T_h}}\Big]^N\\
Z^N_B&=&\sum_{n=1}^{\infty}\Big[2e^{-\frac{\hbar\omega\big(2n+\frac{1}{2}\big)}{k_B T_h}}\Big]^N\\
Z^N_C&=&\sum_{n=1}^{\infty}\Big[2e^{-\frac{\hbar\omega\big(n+\frac{1}{2}\big)}{k_B T_c}}\Big]^N\\
Z^N_D&=&\sum_{n=1}^{\infty}\Big[e^{-\frac{\hbar\omega\big(n+\frac{1}{2}\big)}{k_B T_c}}\Big]^N.
\end{eqnarray}
\end{subequations}
We have not included the over-counting factor as it will be eliminated in the formula for work in the relative partition function. The formula for the work after one cycle of the Stirling-like cycle is given by \cite{thomas2019quantum,zurek1984maxwell,aydiner2021quantum,kim2011quantum} 
\begin{eqnarray}\label{eqn:7}
W^N=k_BT_h \log{\Big[\frac{Z^N_B}{Z^N_A}\Big]}-k_BT_c \log{\Big[\frac{Z^N_C}{Z^N_D}\Big]}.
\end{eqnarray}
We have plotted the work with the particle number $N$ in Fig. (\ref{fig:2}) for a particular frequency of the harmonic potential. Here one can see that the work by the system does not vary significantly with the increase of the number of the non-interacting bosons. 
We also plotted the average work for non-interacting particles in a harmonic potential with respect to the frequency in Fig. (\ref{fig:3}) for $N=1, 2, \text{and}~3$ particles.
	
In Fig. (\ref{fig:3}), the dimensionless work $W/k_B T_c$ is plotted against the frequency of the harmonic potential. As the frequency of the harmonic potential increases at a fixed temperature, the heat capacity of the system vanishes in the high-frequency limit. Therefore, the heat exchanged to lower or raise the temperature during stage ($2$) or stage ($4$) also vanishes. This means that the work by the engine also goes to zero as the frequency of the harmonic potential increases. This means that the engine's efficiency also goes to zero as the frequency of the harmonic potential increases. 
\section{The model}
The QZE for many non-interacting bosons in a quantum harmonic potential can be described as heat supplied to the system from the hot bath and heat absorbed from the system connecting it to the cold bath. In the Stirling-like cycle in Fig. (\ref{fig:1}), the system (a single-particle) is coupled to the hot bath of temperature $T_h$ when the particles are in boxes A and B. Similarly, the system is coupled to the cold bath of temperature $T_c$ when the particles are in boxes C and D \cite{thomas2019quantum,sur2023quantum,aydiner2021quantum}. Therefore from Fig. (\ref{fig:1}), heat is supplied to the system via the processes $(2)$ and $(3)$, and heat is absorbed from the system via the processes $(1)$ and $(4)$ \cite{sur2023quantum}. The processes (1) and (3) are isothermal expansion and compression and processes (2) and (4) are isochoric processes \cite{thomas2019quantum,sur2023quantum,aydiner2021quantum}. The heat exchange between different thermodynamic processes of Fig. (\ref{fig:1}) are given by \cite{thomas2019quantum,kim2011quantum,aydiner2021quantum,sur2023quantum}
\begin{subequations}
\begin{eqnarray}\label{eqn:8}
Q_{AB}&=&U_B-U_A+k_BT_h \log\big[\frac{Z_B}{Z_A}\big]\\
Q_{BC}&=&U_C-U_B\\
Q_{CD}&=&U_D-U_C-k_BT_c \log\big[\frac{Z_C}{Z_D}\big]\\
Q_{DA}&=&U_A-U_D.
\end{eqnarray}
\end{subequations} 
For all reversible cycles $(1)-(4)$ of the Stirling-like cycle, the work to be extracted can be written as \cite{thomas2019quantum,aydiner2021quantum,kim2011quantum,sur2023quantum}
\begin{eqnarray}
W&=&Q_{AB}+Q_{BC}+Q_{CD}+Q_{DA}\nonumber\\
&=&k_BT_h \log\big[\frac{Z_B}{Z_A}\big]-k_BT_c \log\big[\frac{Z_C}{Z_D}\big].\label{eqn:9}
\end{eqnarray}
The heat exchanged between the system of non-interacting bosons and the two baths are given by \cite{sur2023quantum}
\begin{eqnarray}\label{eqn:10}
Q_{hot}&=&Q_{AB}+Q_{DA}\\
\text{and}~~~~~\nonumber\\
Q_{cold}&=&Q_{BC}+Q_{CD}.\label{eqn:11}
\end{eqnarray}
Therefore from the definition of efficiency, the efficiency of the QZE is given by \cite{thomas2019quantum,aydiner2021quantum,kim2011quantum,sur2023quantum}
\begin{eqnarray}
\eta=1+\frac{\text{Heat absorbed}}{\text{Heat Supplied}}
&=&1+\frac{Q_{BC}+Q_{CD}}{Q_{AB}+Q_{DA}}\nonumber\\
=\frac{\text{work}}{\text{Heat supplied}}&=&\frac{W}{Q_{hot}}.\label{eqn:12}
\end{eqnarray}
Now let us consider a particle in a one-dimensional power-law potential of the form 
\begin{eqnarray}\label{eqn:13}
V(x)&=&\alpha |x|^\nu
\end{eqnarray}
where $\nu>0$. The discrete energy spectrum of the particle are given by \cite{mei1998comment,griffiths2018introduction}
\begin{eqnarray}
E_{n, \nu}&=&\alpha\Bigg[\hbar\big(n+1/2\big)\sqrt{\frac{\pi}{2m\alpha}}\frac{\Gamma\big(1/\nu+3/2\big)}{\Gamma\big(1+1/\nu\big)}\Bigg]^{\frac{2\nu}{\nu+2}}\nonumber\\
&=&\Omega(\nu)\Big[n+\frac{1}{2}\Big]^{\frac{2\nu}{\nu+2}}.\label{eqn:14}
\end{eqnarray}
where $\Omega(\nu)$ is a function of mass $m$, frequency of oscillation, and $\nu$. We have calculated the work and the efficiency for the QZE for non-interacting bosons in fractional power-law potential with respect to $\Omega(\nu)$ which is a function of the frequency $\omega$ for a particular $m$ and $\nu$. Eqn. (\ref{eqn:14}) is most accurate for the large quantum numbers \cite{mei1998comment}. For harmonic oscillator with $\nu=2$ and $\alpha=\frac{1}{2}m\omega^2$, the energy eigenvalues are given by $E_{n,2}=(n+1/2)\hbar\omega$ \cite{mei1998comment,griffiths2018introduction,aydiner2021quantum}. We have employed Eqn. (\ref{eqn:14}) for many non-interacting bosons to realize the QZE for fractional powers. Aydiner has numerically demonstrated QZE for a single-particle in fractional power law potential \cite{aydiner2021quantum}. In this article, we have numerically calculated the work and the efficiency of the QZE for non-interacting bosons in fractional power-law potential. In all our calculations we have taken $\alpha=\frac{1}{2}m\omega^2$. We have numerically calculated the work and efficiency as a function of $\Omega(\nu)$.
	
The grand partition function of $N$ non-interacting bosons at temperature $T$ K is given by \cite{pathria2016statistical,landau2013statistical}
\begin{eqnarray}\label{eqn:15}
Z=\prod_{n=1}^{\infty}\Big[\sum_{k=0}^{\infty}e^{-k(E_n-\mu)\beta}\Big]
=\prod_{n=1}^{\infty}\frac{1}{1-e^{-\beta(E_n-\mu)}}
\end{eqnarray}
where $\beta=1/k_BT$ and $\mu$ represents the chemical potential of the system \cite{chandler1988introduction,pathria2016statistical,arovas2013lecture,sur2023quantum,landau2013statistical}. Here we consider a system that has an excess of non-interacting bosons at low temperatures. The condition that the grand partition function converges for a finite temperature $T$, we take the chemical potential to be less than the energy of the ground state \cite{landau2013statistical,sur2023quantum}. The particle number $N$ dependency of the grand partition function can be found in the expression of the chemical potential which can be derived from the constraint \cite{chandler1988introduction,pathria2016statistical,landau2013statistical,sur2023quantum}
\begin{eqnarray}\label{eqn:16}
N&=&\sum_n n_n=\sum_{n=1}^{\infty} \frac{d_n}{e^{\beta(E_n-\mu)}-1}.
\end{eqnarray}
where $d_n$ is the degeneracy of the single-particle occupancy \cite{pathria2016statistical,landau2013statistical}. At a low-temperature, the probability that all the particles can be found in the ground state is very high. Thus the chemical potential in such a situation can be written as 
\begin{eqnarray}\label{eqn:17}
\mu&\approx& E_1-\frac{\log{\big(1+\frac{d_1}{N}\big)}}{\beta}
\end{eqnarray}
where $E_{1}$ represents the ground state energy. As both the ground state energy and the single-particle degeneracy change after inserting the barrier, the chemical potential of the system also changes accordingly \cite{sur2023quantum,pathria2016statistical,kim2011quantum,bengtsson2018quantum,landau2013statistical}. 
	
\begin{figure}
\includegraphics[width=1 \linewidth]{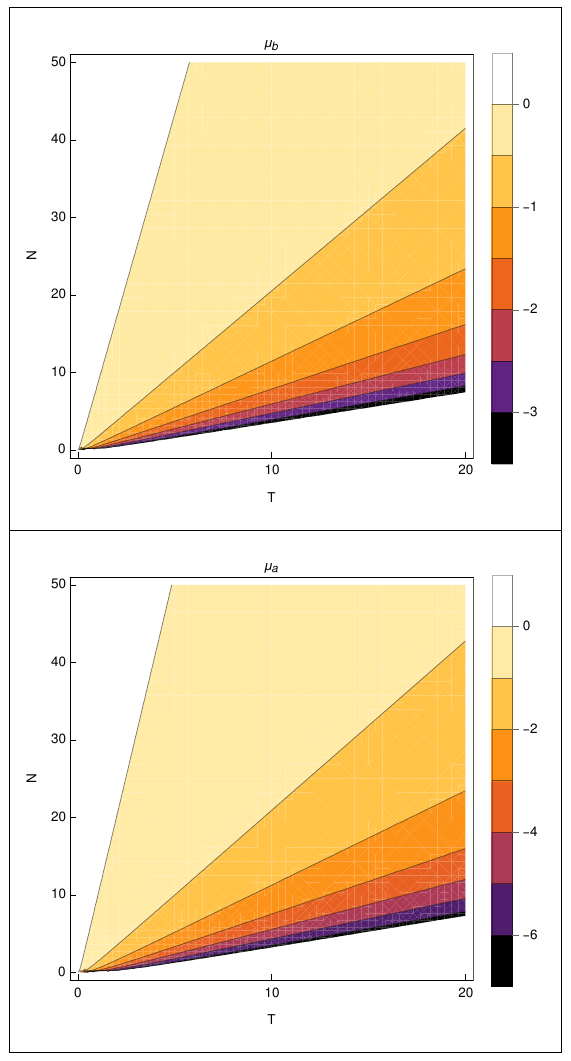}
\caption{Plot of the chemical potentials before and after inserting the barrier i.e. $\mu_b$ and $\mu_a$ for non-interacting bosons Eqn. (\ref{eqn:24}) and Eqn. (\ref{eqn:25}) for $\nu=2$ (harmonic potential) with respect to temperature $T$ and number of bosons $N$ for parameters $\omega=10$ GHz, $\hbar=1.0545\times10^{-34}$J.s, and $k_B=1.3806\times10^{-23}$J$.K^{-1}$. The value of the chemical potentials has been scaled by $10^{23}$ for better visualization.}
\label{fig:4}
\end{figure}

For many non-interacting bosons of equal mass `m' in a power law potential, the relative partition functions of the hot and cold bath at temperatures $T_h$ and $T_c$ can be written as \cite{aydiner2021quantum,li2012revisiting,sur2023quantum}
\begin{eqnarray}\label{eqn:18}
Z[T_h]=\frac{Z_B}{Z_A}=\prod_{n=1}^{\infty}\frac{\Big[1-e^{-\beta_h\big(E_{n,\nu}-\mu_b\big)}\Big]}{\Big[1-e^{-\beta_h\big(E_{2n,\nu}-\mu_a\big)}\Big]}\\
\text{and}~~~~~~~~~~~~~~~~~~~~~~~~~~~~~~~~~~~~~~~~~~~~~~~~~~~~\nonumber\\
Z[T_c]=\frac{Z_C}{Z_D}=\prod_{n=1}^{\infty}\frac{\Big[1-e^{-\beta_c\big(E_{n,\nu}-\mu_b\big]}\Big)}{\Big[1-e^{-\beta_c\big(E_{2n,\nu}-\mu_a\big)}\Big]}\label{eqn:19}
\end{eqnarray}
where $\beta_i=1/k_BT_i$ for $(i=h,~c)$ and $\mu_b$ and $\mu_a$ are the chemical potentials before and after inserting the barrier \cite{sur2023quantum}. The internal energies can be written as \cite{thomas2019quantum,kim2011quantum,aydiner2021quantum}
\begin{eqnarray}\label{eqn:20}
U_{A,B}&=&-\frac{\partial \log{[Z_{A,B}]}}{\partial \beta_h}\\
\text{and}\nonumber\\
U_{C,D}&=&-\frac{\partial \log{[Z_{C,D}]}}{\partial \beta_c}.\label{eqn:21}
\end{eqnarray}
It can be understood from the expression of the efficiency that we only need to calculate the internal energies $U_B$ and $U_D$ to calculate the efficiency of the QZE \cite{thomas2019quantum,aydiner2021quantum}. The internal energies of the boxes B and D for non-interacting bosons in a power law potential are given by \cite{thomas2019quantum,aydiner2021quantum,li2012revisiting,sur2023quantum}
\begin{eqnarray}\label{eqn:22}
U_B&=&\sum_{n=1}^{\infty}2\frac{E_{2n,\nu}-\mu_a}{e^{\beta_h(E_{2n,\nu}-\mu_a)}-1}\\
\text{and}\nonumber\\
U_D&=&\sum_{n=1}^{\infty}\frac{E_{n,\nu}-\mu_b}{e^{\beta_c(E_{n,\nu}-\mu_b)}-1}.\label{eqn:23}
\end{eqnarray}
Factor two in the expression of $U_B$ is due to the double degeneracy of the energy eigenvalues after the barrier is fully inserted \cite{kim2011quantum,bengtsson2018quantum,thomas2019quantum}. The expressions of the chemical potential before and after inserting the barrier for $N$ non-interacting bosons are given by
\begin{eqnarray}\label{eqn:24}
\mu_b&\approx& E^b_1-\frac{\log{\big[1+\frac{1}{N}\big]}}{\beta}\\
\text{and}\nonumber\\
\mu_a&\approx& E^a_1-\frac{\log{\big[1+\frac{2}{N}\big]}}{\beta}\label{eqn:25}
\end{eqnarray}
where $E^b_1$ and $E^a_1$ are the ground state energies of the system before and after inserting the barrier. The chemical potential plays a crucial role in extracting positive work. The chemical potential before and after inserting the barrier must be less than the ground state energies to extract positive work from the heat engine. The chemical potential is a function of the number of particles and of temperature. To ensure a non-zero but less than corresponding ground state energies value of the chemical potentials for a particular range of the frequency $\omega$, we have plotted the chemical potentials with respect to temperature and number of non-interacting bosons in Fig. (\ref{fig:4}).
	
\begin{figure}
\includegraphics[width=1.00 \linewidth]{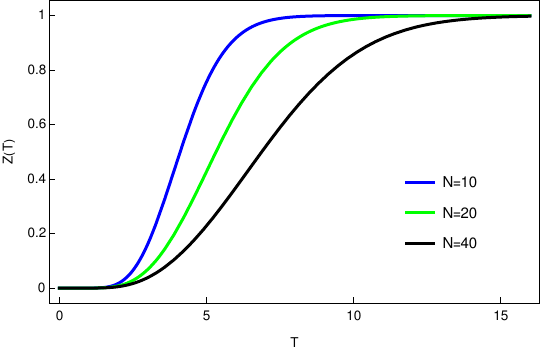}
\caption{Plot of the many-bosons relative partition function $Z(T)$ Eqn. (\ref{eqn:18}) and Eqn. (\ref{eqn:19}) for $\nu=2$ (harmonic potential) with respect to temperature $T$ for different number of bosons. We have used the parameters $\omega=10$ GHz, $\hbar=1.0545\times10^{-34}$J.s, and $k_B=1.3806\times10^{-23}$J$.K^{-1}$.}
\label{fig:5}
\end{figure}
We have also plotted the relative partition function for non-interacting bosons with respect to the temperature for different numbers of particles in Fig. (\ref{fig:5}). The relative partition function for non-interacting bosons is very sensitive to the temperature and number of bosons. For a finite number of bosons, in the high-temperature limit, the relative partition function is unity and thus no work can be extracted from the heat engine. 
\section*{Non-interacting bosons in fractional power law potential}
Depending on the sign of the heat supplied and heat absorbed, many bosons can act like a heat engine or refrigerator \cite{sur2023quantum}. The work for non-interacting bosons in power law potential can be written as  \cite{sur2023quantum,li2012revisiting,thomas2019quantum}
\begin{eqnarray}\label{eqn:26}
W&=&k_BT_h \log{Z[T_h]}-k_BT_c \log{Z[T_c]}\\
\text{and}\nonumber\\
\eta&=&\frac{W}{U_B-U_D+k_BT_h\log{Z[T_h]}}.\label{eqn:26a}
\end{eqnarray}
We begin the analysis considering a single in a harmonic potential. Suppose a barrier is inserted quasi-statically at the position $x=0$. The barrier acts like a delta potential $\lambda(t)\delta(x)$. The time-dependent strength of the delta function barrier can be understood as zero at the beginning of the insertion process and infinity when it is completely inserted \cite{thomas2019quantum,sur2023quantum,zurek1984maxwell,davies2021harmonic,kim2011quantum}. Then the Hamiltonian of the system takes the form
\begin{eqnarray}\label{eqn:27}
H&=&\frac{p^2}{2m}+\frac{1}{2}m\omega^2x^2+\lambda(t)\delta(x).
\end{eqnarray}
After the barrier is fully inserted quasi-statically, the correction in energy eigenvalues for the odd quantum number is zero \cite{viana2011solution,belloni2014infinite,davies2021harmonic}. However, there is a non-zero correction to the energy levels for the even quantum numbers \cite{viana2011solution,belloni2014infinite,davies2021harmonic}. The energy eigenvalues for the even quantum numbers after the barrier is fully inserted can be calculated by graphical method from the equation \cite{goldstein1994supersymmetric,viana2011solution,belloni2014infinite,davies2021harmonic}
\begin{eqnarray}\label{eqn:28}
\frac{\Gamma\Big(\frac{3}{4}-\frac{E}{2\hbar\omega}\Big)}{\Gamma\Big(\frac{1}{4}-\frac{E}{2\hbar\omega}\Big)}&=&-\frac{\lambda^\prime}{2}
\end{eqnarray}
where $\Gamma(.)$ represents the gamma function and $\lambda^\prime=\lambda\sqrt{\frac{m}{\hbar^3\omega}}$. 
\begin{figure}
\includegraphics[width=1.00 \linewidth]{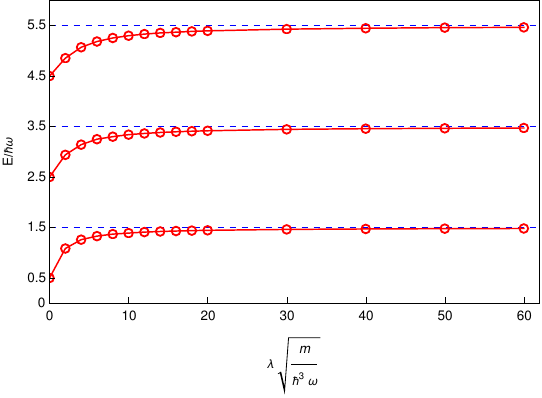}
\caption{Plot of the energy eigenvalues of the odd and even quantum numbers after the barrier is fully inserted~\cite{davies2021harmonic}. The blue dashed lines represent the energy for the odd quantum numbers. The red circles represent the energy of the even quantum numbers and the graphical solution of Eqn. (\ref{eqn:28}). When the barrier is fully inserted i.e. when $\lambda(\infty)\rightarrow\infty$, the energy of the even quantum numbers shifts up and overlaps with the energy of the odd quantum numbers.}
\label{fig:6}
\end{figure}
	
\begin{figure}[H]
\includegraphics[width=1.00 \linewidth]{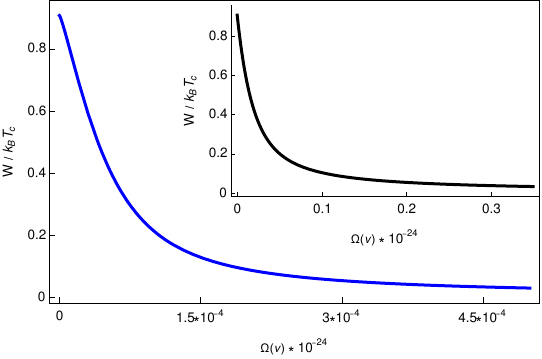}
\caption{Plot of the non-interacting bosons work with respect to $\Omega(\nu)$ for $\nu=1.6$ and follows Eqn. (\ref{eqn:26}). Here we have used the parameters $\hbar=1.0545\times10^{-34}$J.s, and $k_B=1.3806\times10^{-23}$J$.K^{-1}$ $T_h=20$ K, $T_c=10$ K, and $N=20$. The inset shows the same for $\nu=2$ (harmonic potential) and follows Eqn. (\ref{eqn:29}) for the same parameters as mentioned for $\nu=1.6$.}
\label{fig:7}
\end{figure}

For $N$ non-interacting bosons in a harmonic potential ($\nu=2$), the work after one complete cycle of the Stirling-like cycle is given by \cite{bengtsson2018quantum,kim2011quantum,sur2023quantum}
\begin{eqnarray}\label{eqn:29}
W^{har}=k_BT_h\log{\Bigg[ \prod_{n=1}^{\infty}\frac{1-e^{-\beta_h\big(\hbar\omega\big(n+\frac{1}{2}\big)-\mu^{har}_b\big)}}{1-e^{-\beta_h\big(\hbar\omega\big(2n+\frac{1}{2}\big)-\mu^{har}_a\big)}} \Bigg]}\nonumber\\-k_BT_c\log{\Bigg[ \prod_{n=1}^{\infty}\frac{1-e^{-\beta_c\big(\hbar\omega\big(n+\frac{1}{2}\big)-\mu^{har}_b\big)}}{1-e^{-\beta_c\big(\hbar\omega\big(2n+\frac{1}{2}\big)-\mu^{har}_a\big)}} \Bigg]}.
	\end{eqnarray}
Similarly, the internal energies of the boxes B and D for non-interacting bosons in a harmonic potential are given by \cite{li2012revisiting,thomas2019quantum,aydiner2021quantum,sur2023quantum}
\begin{eqnarray}\label{eqn:30}
U^{har}_B&=&\sum_{n=1}^{\infty}2\frac{\hbar\omega\big(2n+\frac{1}{2}\big)-\mu^{har}_a}{e^{\beta_h\big[\hbar\omega(2n+1/2)-\mu^{har}_a\big]}-1}\\
\text{and}\nonumber\\
U^{har}_D&=&\sum_{n=1}^{\infty}\frac{\hbar\omega\big(n+\frac{1}{2}\big)-\mu^{har}_b}{e^{\beta_c\big[\hbar\omega(n+1/2)-\mu^{har}_b\big]}-1}.\label{eqn:31}
\end{eqnarray}
The expression for the chemical potentials before and after inserting the barrier for $N$ non-interacting bosons are given by
\begin{eqnarray}\label{eqn:32}
\mu^{har}_b&\approx& \frac{3\hbar\omega}{2}-k_B T \log{\big[1+\frac{1}{N}\big]}\\
\text{and}\nonumber\\
\mu^{har}_a&\approx& \frac{5\hbar\omega}{2}-k_B T \log{\big[1+\frac{2}{N}\big]}.\label{eqn:33}
\end{eqnarray}
Using all the above equations, we calculate the work and the efficiency for many non-interacting bosons. In Fig. (\ref{fig:7}), we have calculated the work for non-interacting bosons in a fractional power law potential of $\nu=1.6$. 
\begin{figure*}
\begin{minipage}{\textwidth}
\includegraphics[width=1.00 \linewidth]{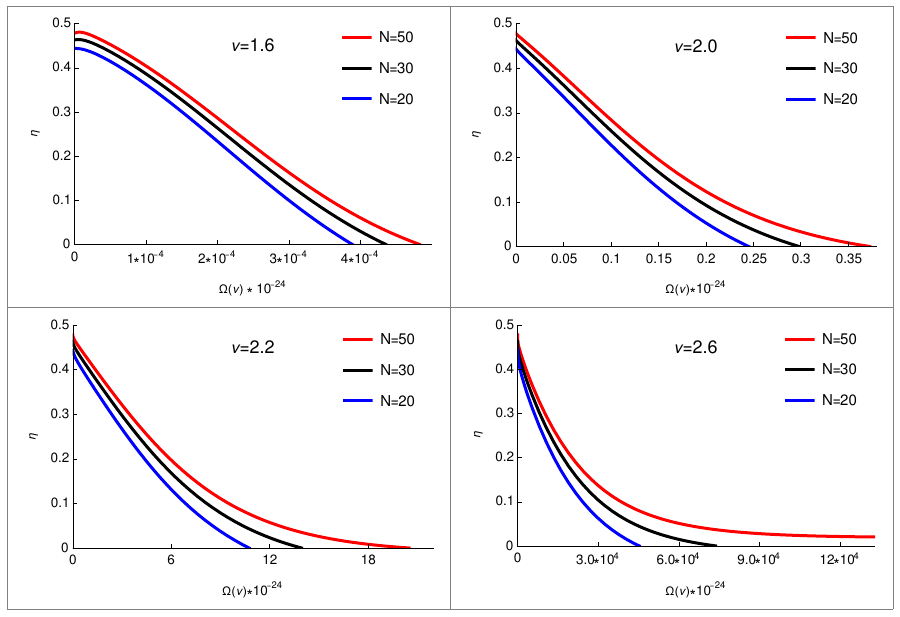}
\caption{Plot of many-boson QZE efficiency with respect to $\Omega(\nu)$ for fractional power law potential for powers $\nu=1.6, 2, 2.2, \text{and}~ 2.6$ and follows Eqn. (\ref{eqn:26a}). Here we have used $m=19.11\times10^{-11}$ Kg, $\hbar=1.0545\times10^{-34}$J.s, and $k_B=1.3806\times10^{-23}$~J$.K^{-1}$, and hot bath and cold bath of temperatures $T_h=2$ K and $T_c=1$ K for different number of particles for different $\nu$.}
\label{fig:8}
\end{minipage}
\end{figure*}
We have calculated the efficiency for different fractional power laws in Fig. (\ref{fig:8}). For comparison, the behavior of the non-interacting bosons efficiency with respect to $\Omega(\nu)$, we have also calculated the efficiency for non-interacting bosons in a harmonic potential. $\lim_{x\to\infty} f(x)$
	
In Fig. (\ref{fig:8}), the efficiency is plotted against the quantity $\Omega(\nu)$ which is a function of the frequency $\omega$. The term $\frac{2\nu}{\nu+2}$ is unity for harmonic potential ($\nu=2$). Therefore the quantity $\Omega(\nu)$ is a linear function of the frequency $\omega$. As the term $\lim_{\nu\to\infty}\frac{2\nu}{\nu+2}\rightarrow2$, for fractional powers $\nu\lesssim2$ or $\nu\gtrsim2$, $\Omega(\nu)$ is of fractional power of greater than or less than one of $\omega$ respectively. When the frequency is zero i.e. $\omega=0$, for negative chemical potentials before and after inserting the barrier, the dimensionless work is positive for $T_c/T_h=0.5$. As the frequency $\omega$ of the power law potential increases at a fixed temperature, the heat capacity of the system decreases and vanishes at higher temperatures. Therefore, the heat exchanged to lower or raise the temperature during stage ($2$) or stage ($4$) also decreases. Thus the efficiency of the QZE also decreases. 
\section*{Quantum Szilard engine for a single-particle in a Morse potential}
Here we consider a single-particle in a Morse potential to realize the QZE. The bound state wavefunction and discrete energy eigenvalues were solved by Morse in 1929 \cite{morse1929diatomic}. The most important feature of using a particle in a Morse potential is its restriction to the number of bound states \cite{morse1929diatomic,simons2003introduction}. The QZE uses the energy quantization of potential well to extract work \cite{kim2011quantum,thomas2019quantum}. Let us consider a single-particle of mass $m$ in a Morse potential of the form
\begin{eqnarray}\label{eqn:34}
V(r)&=&D\Big(1-e^{-\alpha(r-r_e)}\Big)^2
\end{eqnarray}
where $D$ is the disassociation energy or the depth of the potential well \cite{morse1929diatomic,simons2003introduction}. The Morse potential behaves like a harmonic potential in the limit $D\rightarrow\infty$ \cite{morse1929diatomic}. But unlike the harmonic potential energy which is equally spaced between any two adjacent quantum numbers, the energy spacing between any two adjacent quantum numbers decreases as the quantum number increases. The discrete energy eigenvalues of a single-particle in a Morse potential are given by \cite{morse1929diatomic,simons2003introduction}
\begin{eqnarray}\label{eqn:35}
E^{M}_{n,\chi}&=&h\nu\Big(n+\frac{1}{2}\Big)-h\nu\chi\Big(n+\frac{1}{2}\Big)^2
\end{eqnarray}
where the fundamental frequency is given by $\nu=\frac{\alpha}{2\pi}\sqrt{\frac{2D}{m}}$ and the anharmonicity parameter is given by $\chi=\frac{h\nu}{4D}$ \cite{morse1929diatomic,simons2003introduction}. The anharmonicity parameter goes to zero in the limit of infinite disassociation energy i.e. $D\rightarrow\infty$ and the energy eigenvalues behave like harmonic oscillator \cite{morse1929diatomic,simons2003introduction}. The number of bound states of the Morse potential is given by $[n_m]=\Big[\frac{2D}{h\nu}-1\Big]$, where $[.]$ represents the largest integer. As in the low-temperature regime, the quantum effects dominate. The restriction of the maximum number of bound states of the Morse potential can be greatly useful to realize the QZE in the low-temperature regime \cite{chattopadhyay2021bound}.
	
For a single-particle in a quantum harmonic oscillator, the particle can occupy any quantum number ranging from zero to infinity. However, the higher-order terms in the partition function contribute negligibly small in the low-temperature regime \cite{thomas2019quantum}. Thus in the future, it will be interesting to see how the many-particle system (bosons) QZE behaves in the low-temperature limit trapped in a Morse potential. We have calculated the work with respect to the temperature of the hot bath $T_h$ keeping the ratio $\frac{T_c}{T_h}=0.5$ fixed to see how the work of the single-particle QZE in a Morse potential varies with respect to the temperature.
\subsection*{Analysis of the work and the efficiency}
First, we want to clarify that in the high depth limit i.e. $D\rightarrow\infty$, the Morse potential behaves as a quantum harmonic potential \cite{morse1929diatomic,simons2003introduction}. It has also been shown that the even energy levels of the harmonic potential shift up and become double degenerate with the neighboring odd energy levels in the presence of a barrier at $x=0$ in the form delta potential \cite{davies2021harmonic,thomas2019quantum,aydiner2021quantum,belloni2014infinite}. In this article, we have assumed the same for a single-particle in a Morse potential for a large depth of Morse potential on the basis of approximation of wave-function and energy eigenvalues \cite{okock2015matrix}. However, this is not sure for which values quantum number $n$ of the Morse potential, the correction is non-zero finite when a Dirac-delta potential is inserted. To our knowledge, such an analysis has not been done yet by anyone. We kept such analytical analysis as an open problem for the realization of a QZE. What is important in our analysis is that we have employed the energy eigenvalues of the Morse potential and have shown positive work and efficiency in our analysis. However, it is very easy to check that the symmetry of the Morse potential wave function around $x=0$ can be restored numerically for very large values of the depth of the potential i.e. $D\rightarrow\infty$. Then the wavefunction of a single-particle in a Morse potential behaves as a single-particle in a quantum harmonic potential \cite{okock2015matrix}. In this article, we have not attached such numerical analysis. 
	
\begin{figure*}
\includegraphics[width=1.00 \linewidth]{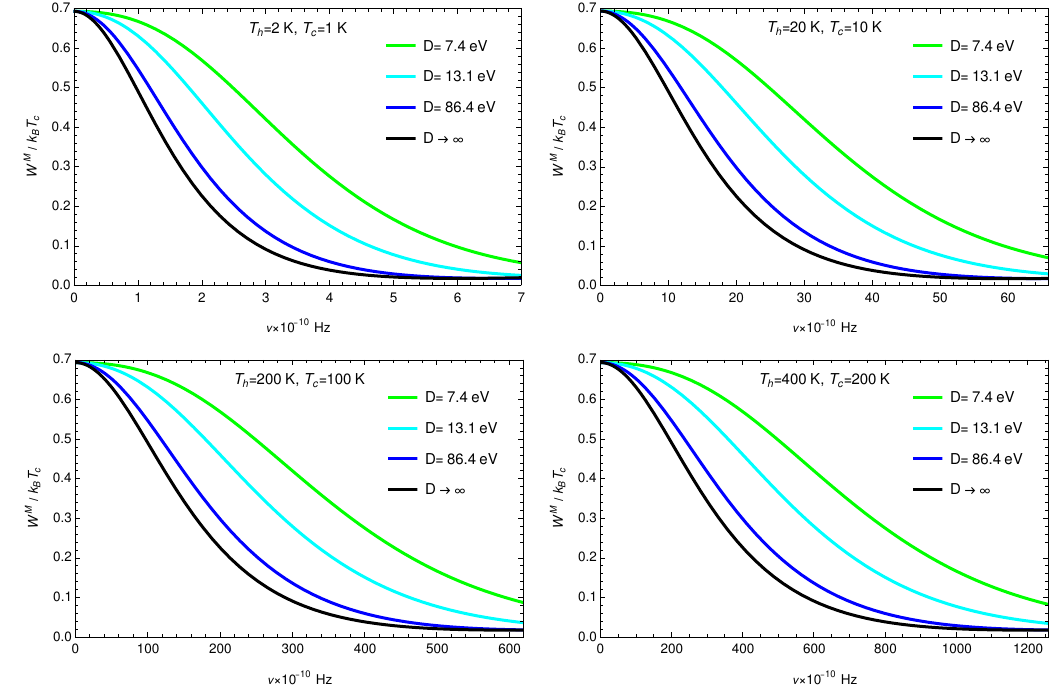}
\caption{Plot of single-particle work with respect to frequency of the Morse potential for different depths of the potential well and four different sets of temperatures. Here we have used $m=1.1$u, $h=6.6260\times10^{-34}$J.s, and $k_B=1.3806\times10^{-23}$J$.K^{-1}$ and follows Eqn. (\ref{eqn:37}).}
\label{fig:10}
\end{figure*}

\begin{figure*}
\includegraphics[width=1.00 \linewidth]{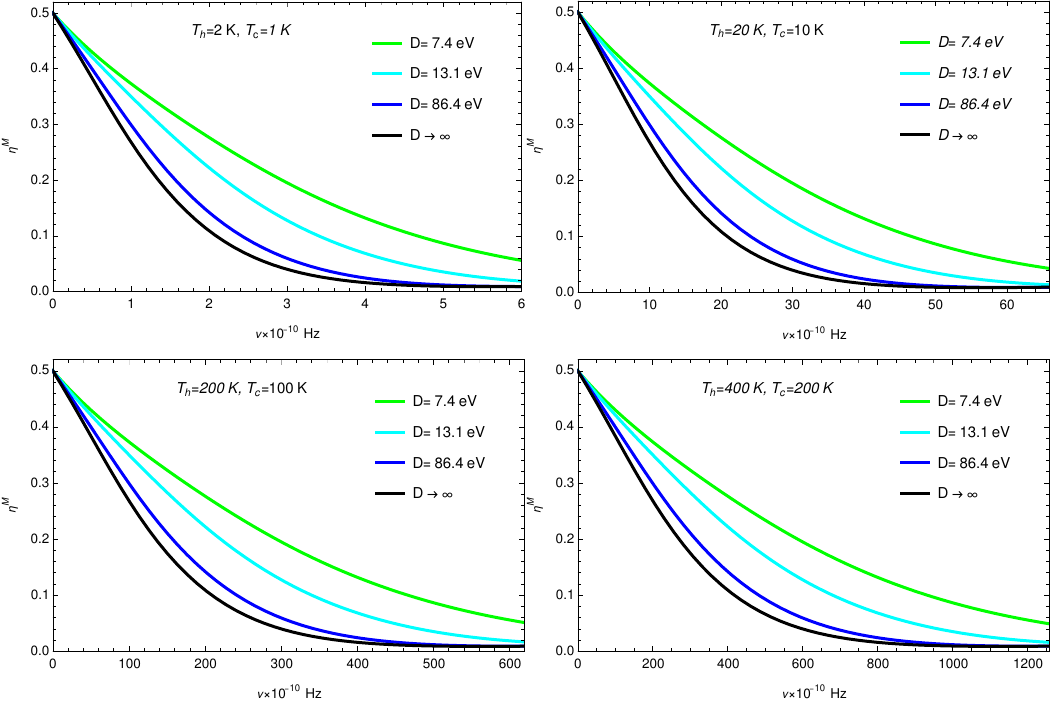}
\caption{Plot of single-particle efficiency with respect to frequency of Morse potential for different depths of the potential well and four different sets of temperatures. Here we have used $m=1.1$u, $h=6.6260\times10^{-34}$J.s, and $k_B=1.3806\times10^{-23}$J$.K^{-1}$ and follows Eqn. (\ref{eqn:38}).}
\label{fig:11}
\end{figure*}

We begin our calculation with the relative partition functions of the four stages and the internal energies of stages 'B' and 'D' for a single-particle in a Morse potential as
\begin{subequations}
\begin{eqnarray}\label{eqn:36}
Z^{M}[T_h]&=&\frac{Z_B[T_h]}{Z_A[T_h]}=\frac{\sum_{n=1}^{[n_m]}2e^{-\beta_h E^{M}_{2n,\chi}}}{\sum_{n=1}^{[n_m]}e^{-\beta_h E^{M}_{n,\chi}}}\\
Z^{M}[T_c]&=&\frac{Z_C[T_c]}{Z_D[T_c]}=\frac{\sum_{n=1}^{[n_m]}2e^{-\beta_c E^{M}_{2n,\chi}}}{\sum_{n=1}^{[n_m]}e^{-\beta_c E^{M}_{n,\chi}}}\\
U^{M}_{B,D}&=&-\frac{\partial}{\partial \beta_{h,c}}\log{\big[Z_{B,D}\big]}.
\end{eqnarray}
\end{subequations}
Therefore, the work and efficiency of a single-particle QZE in a Morse potential after one cycle can be written as
\begin{eqnarray}\label{eqn:37}
W^{M}&=&k_BT_h\log{\big[Z^{M}[T_h]\big]}-k_BT_c\log{\big[Z^{M}[T_c]\big]}~~~\\
\eta^{M}&=&\frac{W^M}{U^{M}_{B}-U^{M}_{D}+k_BT_h Z^{M}[T_h]}.\label{eqn:38}
\end{eqnarray}
\begin{figure}[H]
\includegraphics[width=1.00 \linewidth]{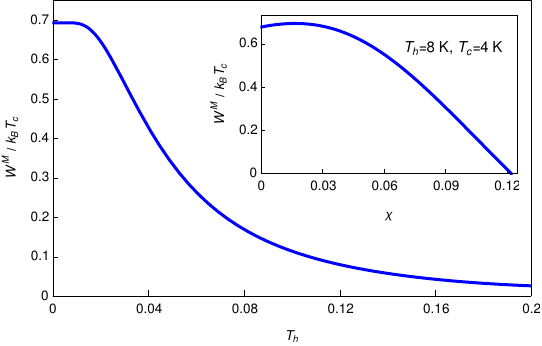}
\caption{Plot of single-particle QZE work in a Morse potential with respect to the temperature of the hot bath $T_h$ keeping the ratio $\frac{T_c}{T_h}=0.5$ fixed for $D=8.7$ eV and follows Eqn. (\ref{eqn:37}). The inset shows the plot of the work with respect to the anharmonicity parameter $\chi$ of the Morse potential and follows Eqn. (\ref{eqn:37}). Here we have used the same parameters with $T_h=8$ K, and $T_c=4$ K respectively. The common parameters for both the plots are $\omega=10$ GHz, $h=6.6260\times10^{-34}$ J.s, and $k_B=1.3806\times10^{-23}$J$.K^{-1}$.}
\label{fig:9}
\end{figure}
At zero temperature, the relative partition functions give two. Thus at zero temperature i.e. when $T\rightarrow0$, the dimensionless work becomes $\log{(2)}$. At a temperature $T$ for which $E^{M}_{n,\chi}$ becomes zero, the relative partition functions become unity and thus give zero work. When the anharmonicity parameter is zero, the relative partition functions $Z^M(T)$ behave like the relative partition for a single-particle in a quantum harmonic potential. The difference comes from the restriction of the upper limit of quantum number $[n_m]$. Then the relative partition function for a finite number of terms becomes $2e^{-\beta_{h,c}\hbar\omega}$. Thus the work then becomes $k_B\big(T_h-T_c\big)\log{(2)}$. When the anharmonicity parameter increases and reaches the value for which $E^{M}_{n,\chi}=0$, the relative partition functions become one giving the work equal to zero. As the efficiency of the engine is proportional to the work, the efficiency varies in the same way as the work. The variation of the work and the efficiency with respect to the frequency $\nu$ follows the same pattern for $E^{M}_{n,\chi}>0$. We have plotted the work and the efficiency with respect to the frequency $\nu$ in Fig. (\ref{fig:10}) and in Fig. (\ref{fig:11}) for different values of the depth of the Morse potential well. We have not considered any low-temperature limit in these cases. We have calculated the work and the efficiency for four different sets of temperatures of the hot bath and of the cold bath for $T_c/T_h=0.5$. 
	
In Fig. (\ref{fig:10}) and in Fig. (\ref{fig:11}), the dimensionless work $W^M/k_B T_c$ and the efficiency $\eta^M$ are plotted against the frequency of the Morse potential. At zero frequency i.e. at $\nu=0$, the relative partition functions give a factor of two. Thus the dimensionless work at zero frequency becomes $\log{(2)}$. When the depth of the Morse potential increases, the anharmonicity parameter becomes smaller, and hence the the second term in energy eigenvalue of the Morse potential becomes less effective. Thus The work by the QZE for a single-particle in the Morse potential behaves similar to the work for a single-particle in a quantum harmonic potential. As the efficiency is proportional to the work, the efficiency varies similarly to the work. The maximum efficiency for a single or non-interacting bosons QZE is given by the Carnot efficiency $\big(1-\frac{T_c}{T_h}\big)$.
\section{Conclusion}
We have numerically obtained results for the work and the efficiency of the QZE for non-interacting bosons in fractional power-law potential. Before applying the Bose-Einstein statistics for non-interacting indistinguishable particles, we have shown that work does not change significantly for $N$ non-interacting indistinguishable particles. We also have calculated the work and the efficiency of the QZE for a single-particle in a Morse potential. In this case, we also have studied the variation of the work with respect to the temperature of the hot bath and the anharmonicity parameter. 
	
We have calculated the work and efficiency with respect to $\Omega(\nu)$ which is a function of the frequency of oscillation for different values of the power law potential including fractional powers. We have taken three different fractional power-law potentials to realize the QZE. We have also considered the case of non-interacting bosons for comparison. We have calculated the efficiency for all four power law potentials at very low temperatures compatible with the Bose-Einstein statistics. We have calculated the efficiency for three sets of non-interacting bosons. We have shown that the efficiency of the QZE increases with the increase in the number of non-interacting bosons. However, the maximum efficiency is restricted by the Carnot efficiency.    
	
We have also calculated the work and the efficiency with respect to the frequency for a single-particle in a Morse potential. We have calculated the work and the efficiency for four sets of temperatures of the hot and cold baths for three different sets of the depth of the Morse potential. We have also compared the work and the efficiency with the harmonic limit i.e. $D\rightarrow\infty$ in all the plots. We have also shown the dependency of the temperature of the hot bath $T_h$ and anharmonicity parameter $\chi$ for the work in this regard. In all the cases our results agree quite well with the limiting values of the work and the efficiency. 
	
Fractional power law potentials do not have closed-form analytical solutions of the Schrodinger equation. As a result, studying quantum systems with these potentials is challenging both analytically and numerically. The relation between thermodynamic information and entropy production leads to the extraction of positive work in the form of a QZE. Fractional power law potentials provide a platform for exploring non-equilibrium quantum thermodynamics. These potentials allow researchers to study the quantum counterparts of classical thermodynamic cycles, such as the Carnot cycle, and investigate quantum fluctuations and correlations in the engine's performance \cite{aydiner2021quantum,sokolov2002fractional,shlesinger1993strange,hilfer2000applications,laskin2000fractional,oldham1974fractional,klages2008anomalous,miller1993introduction}. We study the non-interacting bosons QZE for fractional power-law potentials. This study can be important in many-particle fractional quantum heat engines \cite{aydiner2021quantum}. Fractional power law potentials fall into different universality classes, depending on the exponent of the power law. Each universality class can display unique thermodynamic properties and critical behaviors, making them interesting for understanding phase transitions and quantum critical phenomena \cite{aydiner2021quantum,sokolov2002fractional,shlesinger1993strange,hilfer2000applications,laskin2000fractional,oldham1974fractional,klages2008anomalous,miller1993introduction}. The energy eigenvalues of the fractional power-law potentials for $\nu<2$ or $\nu>2$, deviates the equidistant energy levels of the simple quantum harmonic oscillator. This deviation changes the heat capacity and hence the work and efficiency. Anharmonicity of the Morse potential causes nonlinear energy level transitions, leading to deviations from equidistant energy spacing observed in simple harmonic systems. This alters the quantum states accessible to the engine, affecting the energy absorption and emission during the thermodynamic cycles. Anharmonicity can also influence the quantum fluctuations of the system. These fluctuations can lead to non-classical effects and affect the engine's operation.
	
The energy eigenvalues of the fractional power law potentials derived from the WKB approximation are most appropriate for large quantum numbers.  Thus, our results for the work and the efficiency of the non-interacting bosons QZE would have been better for an exact solution of energy eigenvalues. Fractional power law potentials also introduce anharmonic interactions between particles. However, in this article, we have considered non-interacting particles.  The interaction caused by anharmonicity can lead to deviations from simple harmonic oscillators. These anharmonic effects play a significant role in determining the quantum heat engine's behavior and efficiency. We kept such improvements as an open problem.
	
	
	
\section*{Acknowledgement}
I thank Kabir Ramola for suggesting the problem and for valuable discussions and comments. It is a pleasure to thank Shyamal Biswas and Samir Das for their valuable suggestions and comments.
	
\bibliographystyle{apsrev4-2}
\bibliography{bibfile1.bib}
\clearpage
	
\end{document}